\documentclass[twocolumn,10pt,a4paper]{article}
\usepackage{amsfonts}
\usepackage{amsmath}

\begin{document}

\title{Probabilistic properties of nonextensive thermodynamic}
\author{Franck Jedrzejewski \\
 Commissariat \`{a} l'\'{e}nergie atomique \\
CEA Saclay, INSTN, F-91191 Gif-sur-Yvette, France\\
franck@arthaud.saclay.cea.fr}
\date{}

\maketitle

\begin{abstract}
Based on the Tsallis entropy, the nonextensive thermodynamic properties are
studied as a q-deformation of classical statistical results using only
probabilistic methods and straightforward calculations. It is shown that the
constant in the Tsallis entropy depends on the deformation parameter and
must be redefined to recover the usual thermodynamic relations. The notions
of variance and covariance are generalised. A partial derivative formula of
the entropy is established. It verifies important relations from which most
of the nonextensive thermodynamics relations can be recovered. This leads to
a new proof of a highly nontrivial conclusion that thermodynamic relations
and Maxwell relations in non extensive thermodynamic have the same forms as
those in ordinary nonextensive statistical mechanics. Theoretical results
are applied to ideal gas. The case of fermions and bosons systems with
fractal distributions is also considered. \vskip 3mm \noindent PACS :
05.20.-y, 05.70.-a, 05.30.-d.
\end{abstract}

\section{Introduction}

The purpose of this work is to discuss the thermodynamic relations problem
in Tsallis thermostatistics recently presented in Ref. \cite{Abe0}. It has
been pointed out that $q$-thermostatics could be understood in terms of $q$%
-deformations.\ The Tsallis' entropy proposed by Constantino Tsallis ten
years ago \cite{Tsallis1} has found many applications in various domains of
physics \cite{Tsallis4}, as the turbulence in electron plasmas,
phonon-electron thermalization in ion bombarded solids and flux of solar
neutrinos.\ There are also applications to the mechanisms of anomalous
diffusion \cite{Zanette1}. When systems involve long-range interactions,
statistical mechanics presents serious difficulties due to non-markovianity
of stochastic processes.\ Non-extensive thermodynamics may correctly cover
some of the known anomalies.\ The canonical ensemble of non-extensive
thermodynamics for quantum mechanical systems with finite number of degrees
of freedom has been studied in Ref.\ \cite{Naudts1}. It has been shown that
most of thermodynamic relations are preserved \cite{Abe4}, if introducing
the entropic index $q$, we replace the usual means by $q$-expectations.\
However, invariant relations involve Bolzman-Gibbs constant in the canonical
ensemble and are related to Tsallis constant in the grand canonical
ensemble. The canonical distribution in non-extensive thermodynamic is given
by optimal Lagrange multipliers and is discussed in Ref.\ \cite{Abe3}.
Parameter differentiation of fractional powers of operators has been
developed in Ref. \cite{Raja1,Raja2}. Nonextensive statistical mechanics and
thermodynamics, as a new stream in the research of the foundations of
statistical mechanics, is presented in a recent publication \cite{Springer}.

\section{Tsallis Entropy}

The Tsallis' entropy is defined by the formula%
\begin{equation}
S_{q}=k\int \rho \ln _{q}(\rho ^{-1})\ d\lambda  \label{entropie}
\end{equation}%
where $\rho $\ is the distribution of probability, $\lambda $\ is the
Lebesgue measure on the phase space and $k$ is a constant, called the
Tsallis constant. In this expression the usual logarithmic function is
replaced by the $q$-logarithmic function%
\begin{equation}
\ln _{q}(x)=\frac{x^{1-q}-1}{1-q}
\end{equation}%
which is the inverse function of the $q$-exponential function%
\begin{equation}
e_{q}(x)=(1+(1-q)x)^{1/(1-q)}
\end{equation}%
defined if $(1+(1-q)x)>0$\ and which is prolonged by continuity with $%
e_{q}(x)=0$ otherwise.\ When $q\rightarrow 1$, the $q$-exponential and the $%
q $-logarithmic functions converge to the ordinary exponential and
logarithmic functions.\ The entropy of Tsallis reduces to the Bolzmann-Gibbs
formula in the extensive limit.\ The pseudo-additivity of the $q$%
-logarithmic function%
\begin{equation}
\ln _{q}(xy)=\ln _{q}(x)+\ln _{q}(y)+(1-q)\ln _{q}(x)\ln _{q}(y)
\label{logadd}
\end{equation}%
leads to the nonextensivity of the Tsallis entropy.\ For two independent
subsystems $A$ and $B$ of a total system, in thermal contact with each
other, it has been shown that%
\begin{equation}
S_{q}(\rho _{A}\otimes \rho _{B})=S_{q}(\rho _{A})+S_{q}(\rho
_{B})+(1-q)kS_{q}(\rho _{A})S_{q}(\rho _{B})
\end{equation}%
The following property of the q-logarithmic function%
\begin{equation}
\ln _{q}(\frac{1}{x})=-x^{q-1}\ln _{q}(x)  \label{loginv}
\end{equation}%
gives the usual definition of the Tsallis entropy \cite{Tsallis2}%
\begin{equation}
S_{q}=-k\int \rho ^{q}\ln _{q}(\rho )\ d\lambda
\end{equation}%
which is also defined by the equivalent formula

\begin{equation}
S_{q}=\frac{k}{q-1}\left( 1-\int \rho ^{q}\ d\lambda \right)  \label{entrop}
\end{equation}%
If we define the unnormalized $q$-expectation of an observable $A$ by%
\begin{equation}
\left\lceil A\right\rceil _{q}=\int A\rho ^{q}\ d\lambda
\end{equation}%
and let $c_{q}$ be the constant%
\begin{equation}
c_{q}=\int \rho ^{q}\ d\lambda
\end{equation}%
the $q$-expectation (normalised) could be defined by%
\begin{equation}
\left\langle A\right\rangle _{q}=\frac{\left\lceil A\right\rceil _{q}}{c_{q}}%
=\frac{1}{c_{q}}\int A\rho ^{q}\ d\lambda
\end{equation}%
The problem posed by unnormalized expectation is studied in Ref.\ \cite%
{Pennini}.Introducing the $q$-Gibbs measure (Ref. \cite{Gibbs1}) in the
canonical ensemble 
\begin{equation}
\rho =\frac{e_{q}(-\beta H)}{Z_{q}}
\end{equation}%
where $H$ is the Hamitonian of the system and $\beta $ the Lagrange
multiplier ($1/kT$).\ The expression of the density of probability has been
discussed in Ref.\ \cite{Prato1}.It is shown in Ref.\ \cite{Plastino1} that,
for a canonical distribution of a system in contact with heat bath, the
entropic index $q$ is related to the particle number $N$ and characterizes
the bath. The Tsallis constant $k$ is choosen as 

\begin{equation}
\beta =\frac{Z_{q}^{1-q}}{c_{q}k_{B}T}  \label{beta}
\end{equation}%
where $k_{B}$ is the Boltzamm-Gibbs constant.\ The $q$-partition function $%
Z_{q}$ is given by the normalisation of the probability density 

\begin{equation}
Z_{q}=\int e_{q}(-\beta H)\ d\lambda
\end{equation}%
The measure $\mu $ defined relatively to the Lebesgue measure by $d\mu
=(\rho ^{q}/c_{q})\ d\lambda $ is a measure of probability. In the extensive
limit $q\rightarrow 1$, the $q$-Gibbs measure converges to the ordinary
Boltzmann-Gibbs density. The substitution of the $q$-additivity Eq. (\ref%
{logadd}) in the Tsallis entropy Eq.\ (\ref{entropie}) gives 
\begin{equation}
S_{q}=-k\int (-\beta H+\ln _{q}(Z_{q}^{-1})-(1-q)\beta H\ln
_{q}(Z_{q}^{-1}))\rho ^{q}\ d\lambda
\end{equation}%
and using Eq. (\ref{loginv})%
\begin{equation}
S_{q}=k\int (\beta H+(1-(1-q)\beta H)Z_{q}^{q-1}\ln _{q}(Z_{q}))\rho ^{q}\
d\lambda  \label{seq}
\end{equation}%
we find that the Tsallis entropy

\begin{equation}
S_{q}=k\int (\beta H\rho ^{q}+\ln _{q}(Z_{q})\rho )\ d\lambda
\end{equation}%
leads to the following formula%
\[
S_{q}=k\beta \left\lceil H\right\rceil _{q}+k\ln _{q}(Z_{q}) 
\]%
Introducing the $q$-internal energy%
\begin{equation}
U_{q}=\left\langle H\right\rangle _{q}
\end{equation}%
and the $q$-free energy 
\begin{equation}
F_{q}=\frac{-1}{\beta }\ln _{q}(Z_{q})
\end{equation}%
one finds%
\begin{equation}
F_{q}=c_{q}U_{q}-TS_{q}
\end{equation}%
This relation is the $q$-generalisation of the ordinary equation of $%
S=U/T+kln(Z).$

\section{Mathematical Formalism}

In this section, we introduce five important relations for the nonextensive
mathematical formalism.\ The generalised internal energy $U_{q}$ defined as
the $q$-expectation of the Hamitonian is given by%
\begin{equation}
U_{q}=-\frac{1}{c_{q}}\frac{\partial \ln _{q}(Z_{q})}{\partial \beta }
\label{uq}
\end{equation}%
The proof of this equation is easy to compute.\ 
\begin{eqnarray}
-\frac{\partial \ln _{q}(Z_{q})}{\partial \beta } &=&-\frac{\partial \ln
_{q}(Z_{q})}{\partial Z_{q}}\frac{\partial Z_{q}}{\partial \beta }=\frac{-1}{%
Z_{q}^{q}}\frac{\partial Z_{q}}{\partial \beta } \\
&=&\frac{-1}{Z_{q}^{q}}\frac{\partial }{\partial \beta }\int e_{q}(-\beta
H)\ d\lambda
\end{eqnarray}%
Using the following expression of the derivative of the $q$-exponential
function with respect to $\beta $%
\begin{equation}
\frac{\partial e_{q}(-\beta H)}{\partial \beta }=-He_{q}^{q}(-\beta H)
\end{equation}%
leads to%
\begin{eqnarray}
-\frac{\partial \ln _{q}(Z_{q})}{\partial \beta } &=&\int H\frac{%
e_{q}^{q}(-\beta H)}{Z_{q}^{q}}\ d\lambda \\
&=&\int H\rho ^{q}\ d\lambda =\left\lceil H\right\rceil _{q}=c_{q}U_{q}
\end{eqnarray}%
Defining the $q$-movariance of two operators $A$ and $B$ by%
\begin{equation}
Mov_{q}(A,B)=\left\lceil AB\right\rceil _{2q-1}-\left\lceil A\right\rceil
_{q}\left\lceil B\right\rceil _{q}
\end{equation}%
the equation (Eq.\ \ref{M01})\ gives the $q$-movariance of $A$ and $H$ in
function of the derivative of the $q$-mean of $A$%
\begin{equation}
\frac{\partial \left\lceil A\right\rceil _{q}}{\partial \beta }-\left\lceil 
\frac{\partial A}{\partial \beta }\right\rceil _{q}=-qZ_{q}^{q-1}Mov_{q}(A,H)
\label{M01}
\end{equation}%
The movariance converges to the usual covariance in the extensive limit $%
q\rightarrow 1$. Proof of the equation (Eq.\ \ref{M01}) is 
\begin{eqnarray}
\frac{\partial \left\lceil A\right\rceil _{q}}{\partial \beta } &=&\int 
\frac{\partial A}{\partial \beta }\rho ^{q}\ d\lambda +\int A\frac{\partial
\rho ^{q}}{\partial \rho }\frac{\partial \rho }{\partial \beta }\ d\lambda
\label{s01} \\
&=&\left\lceil \frac{\partial A}{\partial \beta }\right\rceil _{q}+q\int
A\rho ^{q-1}\frac{\partial }{\partial \beta }\left( \frac{e_{q}(-\beta H)}{%
Z_{q}}\right) \ d\lambda
\end{eqnarray}%
The derivative of the probability density with respect to the Lagrange
multiplier is computed as%
\begin{eqnarray}
\frac{\partial \rho }{\partial \beta } &=&\frac{1}{Z_{q}}\frac{\partial
e_{q}(-\beta H)}{\partial \beta }-\frac{e_{q}(-\beta H)}{Z_{q}^{2}}\frac{%
\partial Z_{q}}{\partial \beta } \\
&=&\frac{-H}{Z_{q}}e_{q}^{q}(-\beta H)+\frac{e_{q}(-\beta H)}{Z_{q}^{2}}\int
He_{q}^{q}(-\beta H)\ d\lambda
\end{eqnarray}%
Substituing the previous equation in Eq. (\ref{s01}) and using that $Z_{q}$
is a constant leads to%
\begin{eqnarray}
\Delta _{q,\beta }A &=&\frac{\partial \left\lceil A\right\rceil _{q}}{%
\partial \beta }-\left\lceil \frac{\partial A}{\partial \beta }\right\rceil
_{q}  \nonumber \\
&=&-qZ_{q}^{q-1}\int AH\rho ^{2q-1}\ d\lambda +\cdots \\
&&+qZ_{q}^{q-1}\int A\rho ^{q}\left( \int H\rho ^{q}\ d\lambda \right) \
d\lambda \\
&=&-qZ_{q}^{q-1}(\left\lceil AH\right\rceil _{2q-1}-\left\lceil
A\right\rceil _{q}\left\lceil H\right\rceil _{q}) \\
&=&-qZ_{q}^{q-1}Mov_{q}(A,H)
\end{eqnarray}%
If the operator $A$ does not depend explicitly of $\beta ,$\ we find that
the $q$-movariance is proportionnal to the derivative of the $q$-expectation
of $A$ with respect to $\beta .$\ We define the $q$-mariance by%
\begin{equation}
Mar_{q}(A)=\left\lceil A^{2}\right\rceil _{2q-1}-\left\lceil A\right\rceil
_{q}^{2}
\end{equation}%
This function converges to the usual variance when $q$ tends to $1$. Using
the equation of the generalised energy Eq. (\ref{uq}), we see that the $q$%
-mariance of the hamiltonian is given by%
\begin{equation}
Mar_{q}(H)=\frac{1}{qZ_{q}^{q-1}}\frac{\partial ^{2}\ln _{q}(Z_{q})}{%
\partial \beta ^{2}}
\end{equation}%
Applying Cauchy-Schwarz inequality%
\begin{equation}
\left( \int fg\ d\lambda \right) ^{2}\leq \left( \int f^{2}\ d\lambda
\right) \left( \int g^{2}\ d\lambda \right)
\end{equation}%
with $f=A\rho ^{q-1/2}Z_{q}^{q-1}$ and $g=\rho ^{1/2}$ proves that $%
Mar_{q}(A)$ is always positive.\ In the extensive limit when $q$ tends to $1$%
, the variance of the hamitonian $Var(H)=\partial ^{2}\ln (Z)/\partial \beta
^{2}$ is positive.\ Thus the internal energy as a function of the Lagrange
multiplier $\beta $\ is a convex function.

\bigskip

We suppose now that the Hamiltonian $H$ and the probability density function 
$\rho $\ depend on a parameter $\alpha $ distinct of $\beta $.%
\begin{equation}
\frac{\partial \ln _{q}(Z_{q})}{\partial \alpha }=\beta \left\lceil \frac{%
\partial H}{\partial \alpha }\right\rceil _{q}
\end{equation}%
Remark that this equation is also valid for $\alpha $ equal$\ \beta $ (by
Eq.\ (\ref{uq})). Proof of this equation is easy to compute.%
\begin{equation}
-\frac{\partial \ln _{q}(Z_{q})}{\partial \alpha }=\frac{-1}{Z_{q}^{q}}\frac{%
\partial Z_{q}}{\partial \alpha }=\frac{-1}{Z_{q}^{q}}\frac{\partial }{%
\partial \alpha }\int e_{q}(-\beta H)\ d\lambda
\end{equation}%
Using the expression of the derivative of the $q$-exponential function with
respect to $\alpha $%
\begin{equation}
\frac{\partial e_{q}(-\beta H)}{\partial \alpha }=-\beta \frac{\partial H}{%
\partial \alpha }e_{q}^{q}(-\beta H)  \label{deriveq}
\end{equation}%
leads to%
\begin{equation}
-\frac{\partial \ln _{q}(Z_{q})}{\partial \beta }=\beta \int \frac{\partial H%
}{\partial \alpha }\rho ^{q}\ d\lambda =\beta \left\lceil \frac{\partial H}{%
\partial \alpha }\right\rceil _{q}
\end{equation}%
The next fundamental equation gives the expression of the movariance of an
observable and the derivation of the hamiltonian with respect to a given
parameter 
\begin{equation}
\frac{\partial \left\lceil A\right\rceil _{q}}{\partial \alpha }-\left\lceil 
\frac{\partial A}{\partial \alpha }\right\rceil _{q}=-q\beta
Z_{q}^{q-1}Mov_{q}(A,\frac{\partial H}{\partial \alpha })  \label{eq04}
\end{equation}%
The proof of this equation is 
\begin{eqnarray}
\frac{\partial \left\lceil A\right\rceil _{q}}{\partial \alpha } &=&\int 
\frac{\partial A}{\partial \alpha }\rho ^{q}\ d\lambda +\int A\frac{\partial
\rho ^{q}}{\partial \rho }\frac{\partial \rho }{\partial \alpha }\ d\lambda
\\
&=&\left\lceil \frac{\partial A}{\partial \beta }\right\rceil _{q}+q\int
A\rho ^{q-1}\frac{\partial }{\partial \alpha }\left( \frac{e_{q}(-\beta H)}{%
Z_{q}}\right) \ d\lambda
\end{eqnarray}%
The derivative of the partition function with respect to the parameter is
given by%
\begin{equation}
\frac{\partial Z_{q}}{\partial \alpha }=-\beta \int \frac{\partial H}{%
\partial \alpha }e_{q}^{q}(-\beta H)\ d\lambda  \label{derivzq}
\end{equation}%
Let $\Delta _{q,\alpha }A$\ be the quantity 
\begin{equation}
\Delta _{q,\alpha }A=\frac{\partial \left\lceil A\right\rceil _{q}}{\partial
\alpha }-\left\lceil \frac{\partial A}{\partial \alpha }\right\rceil _{q}
\end{equation}%
Substituing Eq.\ (\ref{deriveq}) and (\ref{derivzq}) in the previous
equation and using that $Z_{q}$ is a constant, we get%
\begin{eqnarray}
\Delta _{q,\alpha }A &=&q\int A\rho ^{q-1}\frac{\partial }{\partial \alpha }%
\left( \frac{e_{q}(-\beta H)}{Z_{q}}\right) \ d\lambda \\
&=&-qZ_{q}^{q-1}\beta \int A\frac{\partial H}{\partial \alpha }\rho ^{2q-1}\
d\lambda +\cdots \\
&&+qZ_{q}^{q-1}\beta \int A\rho ^{q}\left( \int \frac{\partial H}{\partial
\alpha }\rho ^{q}\ d\lambda \right) \ d\lambda \\
&=&-qZ_{q}^{q-1}\left( \left\lceil A\frac{\partial H}{\partial \alpha }%
\right\rceil _{2q-1}-\left\lceil A\right\rceil _{q}\left\lceil \frac{%
\partial H}{\partial \alpha }\right\rceil _{q}\right) \\
&=&-qZ_{q}^{q-1}Mov_{q}(A,\frac{\partial H}{\partial \alpha })
\end{eqnarray}%
The following equation characterizes the derivation of the Tsallis entropy
with respect to the $\alpha $\ parameter%
\begin{eqnarray}
\frac{\partial S_{q}}{\partial \alpha } &=&-k\beta \left( \frac{\partial
\left\lceil H\right\rceil _{q}}{\partial \alpha }-\left\lceil \frac{\partial
H}{\partial \alpha }\right\rceil _{q}\right) \\
&=&qk\beta ^{2}Z_{q}^{q-1}Mov_{q}(H,\frac{\partial H}{\partial \alpha })
\end{eqnarray}%
In the extensive limit when $q$ tends to $1$, this equation shows that the
derivation of the usual entropy is proportional to the variance of the
hamiltonian%
\begin{equation}
\frac{\partial S}{\partial \alpha }=\frac{1}{T}Var(H)=\frac{1}{T}\left( 
\frac{\partial \left\langle H\right\rangle }{\partial V}-\left\langle \frac{%
\partial H}{\partial V}\right\rangle \right)
\end{equation}%
If the parameter $\alpha $\ is the volume $V$ and the pressure defined by $%
P=-\left\langle \partial H/\partial V\right\rangle ,$ this equation is the
usual thermodynamic relation $TdS=dU+PdV$. If we define the $q$-pressure by%
\begin{equation}
P_{q}=-\left\langle \frac{\partial H}{\partial V}\right\rangle _{q}=-\frac{1%
}{\beta c_{q}}\frac{\partial \ln _{q}(Z_{q})}{\partial V}
\end{equation}%
and write the derivative of the Tsallis's entropy with respect to the volume%
\begin{eqnarray}
\frac{\partial S_{q}}{\partial V} &=&-k\beta c_{q}\left( \frac{\partial
\left\langle H\right\rangle _{q}}{\partial V}-\left\langle \frac{\partial H}{%
\partial V}\right\rangle _{q}\right) \\
&=&-k\beta c_{q}\left( \frac{\partial U_{q}}{\partial V}+P_{q}\right)
\end{eqnarray}%
leads to the ordinary equation 
\begin{equation}
dU_{q}=-c_{q}TdS_{q}-P_{q}dV
\end{equation}%
Define the following q-quantities%
\begin{equation}
\frac{\partial Q_{q}}{\partial V}=\frac{\partial \left\langle H\right\rangle
_{q}}{\partial V}-\left\langle \frac{\partial H}{\partial V}\right\rangle
_{q}
\end{equation}%
\begin{equation}
\frac{\partial W_{q}}{\partial V}=-P_{q}=\left\langle \frac{\partial H}{%
\partial V}\right\rangle _{q}
\end{equation}%
leads to the usual equation in the q-world%
\begin{equation}
dU_{q}=dQ_{q}+dW_{q}=dQ_{q}-P_{q}dV
\end{equation}%
Derivating Eq.\ (\ref{entrop}) with respect to parameter $\alpha $%
\begin{equation}
\frac{\partial S_{q}}{\partial \alpha }=\frac{k}{1-q}\frac{\partial c_{q}}{%
\partial \alpha }
\end{equation}%
Substituing the following equation%
\begin{equation}
c_{q}=1+\frac{1-q}{k}S_{q}
\end{equation}%
and taking $A=1$ in Eq.\ (\ref{eq04}) leads to the differential equation
fillows by the Tsallis entropy.%
\begin{eqnarray}
&&\frac{\partial S_{q}}{\partial \alpha }-q\beta Z_{q}^{q-1}\frac{\partial
\left\lceil H\right\rceil _{q}}{\partial \alpha }S_{q} \\
&=&\frac{q}{q-1}k\beta Z_{q}^{q-1}\left( \frac{\partial \left\lceil
H\right\rceil _{2q-1}}{\partial \alpha }-\frac{\partial \left\lceil
H\right\rceil _{q}}{\partial \alpha }\right)
\end{eqnarray}

\section{Ideal Gas}

The classical ideal gas in D-dimensional space has been exhaustively
discussed in Ref.\ \cite{Abe1}. Correlations were analyzed in Ref.\cite{Abe2}%
\ . We summarize some results in the case $0<q<1$ and show by
straightforward calculation how the partition function is computed.\ The
hamiltonian is%
\begin{equation}
H=%
\mathrel{\mathop{\stackrel{N}{\sum }}\limits_{i=1}}%
\frac{p_{i}^{2}}{2m}
\end{equation}%
where $m$ is the particle mass, $N$\ the particle number and $p_{i}$ the
momentum of the ith particle. The partition function is given by%
\begin{equation}
Z_{q}=\int_{\Omega }(1-a(p_{1}^{2}+...+p_{N}^{2}))^{\frac{1}{1-q}}\ \frac{%
d^{D}p_{i}d^{D}r_{i}}{N!h^{DN}}
\end{equation}%
where $a$ is the positive constant $a=(1-q)\beta /2m$ and $\Omega $\ is
domain of integration such that $1-a(p_{1}^{2}+...+p_{N}^{2})>0.$%
\begin{equation}
Z_{q}=\frac{V^{N}}{N!h^{DN}}\int_{\Omega }(1-a(p_{1}^{2}+...+p_{N}^{2}))^{%
\frac{1}{1-q}}\ d^{D}p_{i}
\end{equation}%
and%
\begin{equation}
Z_{q}=\frac{V^{N}}{N!h^{DN}}\frac{2\pi ^{DN/2}}{\Gamma (DN/2)}\int_{0}^{1/%
\sqrt{a}}(1-ar^{2})^{\frac{1}{1-q}}r^{\frac{DN}{2}-1}\ dr
\end{equation}%
Let $x$ be $ar^{2}.$ Using the fact that the beta distribution is a density
of probability for two parameters $u$ and $v$ greatest or equal 1, that is%
\begin{equation}
\frac{\Gamma (u+v)}{\Gamma (u)\Gamma (v)}\int_{0}^{1}x^{u-1}(1-x)^{v-1}dx=1
\end{equation}%
After a simple calculation, we find the following expression of the $q$%
-partition function 
\begin{equation}
Z_{q}=\frac{V^{N}}{N!}\left( \frac{2\pi m}{(1-q)\beta h^{2}}\right) ^{DN/2}%
\frac{\Gamma (\frac{2-q}{1-q})}{\Gamma (\frac{2-q}{1-q}+\frac{DN}{2})}
\end{equation}%
The same calculation for the constant $c_{q}$ leads to%
\begin{equation}
c_{q}=Z_{q}^{1-q}\frac{\Gamma (\frac{1}{1-q})}{\Gamma (\frac{1}{1-q}+\frac{DN%
}{2})}\frac{\Gamma (\frac{2-q}{1-q}+\frac{DN}{2})}{\Gamma (\frac{2-q}{1-q})}
\end{equation}%
Using the following property of the gamma function 
\begin{equation}
\Gamma (z+1)=z\Gamma (z)
\end{equation}%
we find%
\begin{equation}
c_{q}=Z_{q}^{1-q}(1+(1-q)\frac{DN}{2})
\end{equation}%
The generalised energy given by Eq. (\ref{uq}) is computed in the same way%
\begin{equation}
U_{q}=\frac{DN}{2\beta }\frac{Z_{q}^{1-q}}{c_{q}}
\end{equation}%
Using the expression of the Lagrange mutiplier given by Eq. (\ref{s01}), we
obtain the following expression of the internal energy which is independant
of the entropic index.%
\begin{equation}
U_{q}=\frac{DN}{2}k_{B}T
\end{equation}%
The Tsallis and Boltzmann-Gibbs constants are related to the equation%
\begin{equation}
k=c_{q}Z_{q}^{q-1}k_{B}=\left( 1-(1-q)\frac{DN}{2}\right) k_{B}
\end{equation}%
\ The equation of state%
\begin{equation}
P_{q}=\frac{1}{\beta c_{q}Z_{q}^{q}}\frac{\partial Z_{q}}{\partial V}=\frac{N%
}{V}\frac{Z_{q}^{1-q}}{\beta c_{q}}
\end{equation}%
leads to the usual expression%
\begin{equation}
P_{q}V=Nk_{B}T
\end{equation}%
In the extensive limit when $q$ tends to $1$, we recover the equation of
ideal gas. Remark that if we replace the hamiltonian by the same expression
minus the generalized energy, the results are unchanged \cite{Rajagopal}.\
The $q$-mariance of the hamitonian 
\begin{equation}
Marq(H)=\frac{1}{q}k_{B}TU_{q}
\end{equation}%
converges in the extensive limit to the variance of the hamiltonian $%
DNk_{B}^{2}T^{2}/2$. Simple calculations give the movariance of the $q$%
-pressure%
\begin{equation}
Mov_{q}(H,P_{q})=\frac{1-q}{q}\frac{DN}{2\beta }P_{q}c_{q}Z_{q}^{1-q}
\end{equation}%
and 
\begin{equation}
Mov_{q}(H,\frac{\partial H}{\partial V})=\frac{-Nk}{qVk_{B}\beta ^{2}}%
Z_{q}^{2-2q}
\end{equation}%
and the ratio%
\begin{equation}
\frac{Mov_{q}(H,P_{q})}{Mov_{q}(H,\frac{\partial H}{\partial V})}=(q-1)\frac{%
k}{k_{B}}\frac{DN}{2}
\end{equation}

\section{Quantum Free Particle}

In this section, a free particle confined in a 3-dimensional box is
considered.\ Denoting $L_{1}$, $L_{2}$, $L_{3}$ the sides of the box, it is
well known that the energy eigenvalues are given by%
\begin{equation}
E=\frac{h^{2}}{8m}\left( \frac{n_{1}^{2}}{L_{1}^{2}}+\frac{n_{2}^{2}}{%
L_{2}^{2}}+\frac{n_{3}^{2}}{L_{3}^{2}}\right)
\end{equation}%
where the quantum numbers are non-negative integers $n_{i}=0,1,2$, ...The
partition function is given by%
\begin{equation}
Z_{q}=%
\mathrel{\mathop{\sum }\limits_{n_{1},n_{2},n_{3}}}%
e_{q}(-\beta E)=%
\mathrel{\mathop{\sum }\limits_{n_{1},n_{2},n_{3}}}%
\left( 1-(1-q)E\right) ^{\frac{1}{1-q}}
\end{equation}%
Let $\lambda _{T}$ be the translation temperature%
\begin{equation}
\lambda _{T}=\frac{h^{2}}{8mkL^{2}}
\end{equation}%
where $L\sim L_{i}$ is the length of a side.\ For sufficiently large $T\gg
\lambda _{T}$, the sum over $n_{i}$ can be replaced by an integral%
\begin{equation}
Z_{q}=\frac{1}{8}\int \left( 1-(1-q)\beta E\right) ^{\frac{1}{1-q}}{\Large 1}%
_{\Omega }(n_{i})dn_{1}dn_{2}dn_{3}
\end{equation}%
where $\Omega =\{1-(1-q)\beta E>0\}$ and $1_{A}(x)=1$ if $x\in A$ and 0
otherwise. For a volume $V=L_{1}L_{2}L_{3}$ and a parameter $b=\beta
(1-q)h^{2}/8m$ the change of variables 
\begin{equation}
Z_{q}=\frac{V}{8}\frac{2\pi ^{3/2}}{\Gamma (3/2)}\int_{0}^{1/\sqrt{b}%
}(1-br^{2})^{\frac{1}{1-q}}r^{2}dr
\end{equation}%
could be written as a beta integral%
\begin{equation}
Z_{q}=\frac{V}{8}\frac{\pi ^{3/2}}{b^{3/2}\Gamma (3/2)}\int_{0}^{1}(1-x)^{%
\frac{1}{1-q}}x^{\frac{1}{2}}dr
\end{equation}%
thus%
\begin{equation}
Z_{q}=V\left( \frac{2\pi }{\beta (1-q)h^{2}}\right) ^{3/2}\frac{\Gamma (%
\frac{2-q}{1-q})}{\Gamma (\frac{2-q}{1-q}+\frac{3}{2})}
\end{equation}%
this is the classical formula with $N=1$ and $D=3$.

\section{Fermions and Bosons Distributions}

Thermodynamics on quantum groups and thermodynamics of boson and fermion
systems with fractal distributions have been studied recently \cite%
{Ubriaco1,Ubriaco2,Raja3}.\ In the grand canonical ensemble, the grand
partition function is the sum under the states $j$ 
\begin{equation}
\Theta _{q}=%
\mathrel{\mathop{\sum }\limits_{j}}%
\left( 1-\beta (1-q)(E_{j}-\mu )\right) ^{\frac{1}{1-q}}
\end{equation}%
where $\mu $\ is the chemical potential.\ The density of states in the phase
space is given by%
\begin{equation}
D(E)=\frac{\partial }{\partial E}\int \frac{d^{3}rd^{3}p}{h^{3}}=\frac{2\pi V%
}{h^{3}}(2m)^{3/2}\sqrt{E}
\end{equation}%
In the thermodynamic limit, the sum is replaced by a continous integration,
thus the average is defined by%
\begin{equation}
\left\langle A\right\rangle _{q}=%
\mathrel{\mathop{\sum }\limits_{j}}%
A_{j}p_{j}^{q}\simeq \int_{0}^{\infty }AD(E)f^{q}(E)dE
\end{equation}%
where $f$ is the function%
\begin{equation}
f(E)=\frac{1}{\left( 1-\beta (1-q)(E-\mu )\right) ^{\frac{1}{q-1}}}{\Large 1}%
_{\Omega }
\end{equation}%
The change of variable $E=x^{2}/\beta (q-1)$ gives%
\begin{equation}
\left\langle A\right\rangle _{q}=\frac{4\pi V}{h^{3}}\left( \frac{2mkT}{q-1}%
\right) ^{3/2}\int_{0}^{\infty }A^{q}x^{2}f(x)dx
\end{equation}%
The average total number of particles%
\begin{equation}
\overline{N}=%
\mathrel{\mathop{\sum }\limits_{j}}%
\frac{1}{\left( 1-\beta (1-q)(E_{j}-\mu )\right) ^{\frac{1}{q-1}}}
\end{equation}%
in the thermodynamic limit%
\begin{equation}
\overline{N}\simeq \int_{0}^{\infty }D(E)f(E)dE
\end{equation}%
is calculated by 
\begin{equation}
\overline{N}=K\int_{a}^{\infty }\frac{x^{2}}{\left( 1+x^{2}-\beta (q-1)\mu
\right) ^{\frac{1}{1-q}}}dx
\end{equation}%
where $a^{2}=-1+\beta (q-1)\mu $ and%
\begin{equation}
K=\frac{4\pi V}{h^{3}}\left( \frac{2mkT}{q-1}\right) ^{3/2}
\end{equation}%
This integral could be computed as a beta integral if $1\leq q\leq 7/5$%
\begin{equation}
\overline{N}=\frac{4\pi V}{3(q-1)h^{3}}\left( \frac{2mkT}{q-1}\right)
^{3/2}a^{-2u}\frac{\Gamma (u)\Gamma (v)}{\Gamma (u+v)}
\end{equation}%
where $u=(5-3q)/2(q-1)$ and $v=1/(1-q)$. Same calculation could be done for
the generalised energy in the thermodynamic limit%
\begin{equation}
U_{q}=\frac{2\pi V}{h^{3}}\int_{0}^{\infty }(2mE)^{3/2}f^{q}(E)dE
\end{equation}%
after simple manipulation%
\begin{equation}
U_{q}=\frac{4\pi V}{2\beta (q-1)h^{3}}\left( \frac{2mkT}{q-1}\right)
^{3/2}a^{-2u}\frac{\Gamma (u)\Gamma (v)}{\Gamma (u+v)}
\end{equation}%
The average energy per particle 
\begin{equation}
U_{q}=\frac{3}{2}\overline{N}kT
\end{equation}%
It is worth noting that the generalized energy is in this equation related
to the Tsallis constant (and not to the Boltzmann-Gibbs constant).

\section{Conclusions}

In this paper, we studied the probabilistic properties within the framework
of Tsallis thermodynamics.\ It has been shown that movariance and mariance
are suitable functions for generalized covariance and variance in the
non-extensive thermodynamic context.\ The invariance of thermodynamic
relations is established if the Tsallis constant depend on the entropic
parameter.\ However, in the canonical ensemble, as illustrated on the ideal
gas, the equation of state as the same form as the ordinary equation in the
non-extensive formalism with the Bolzmann-Gibbs constant. In the grand
canonical ensemble, the dependance of the $q$-average internal energy is
related to the Tsallis constant.\ In both cases, we recover the usual
equations in the extensive limit.

\section*{Acknowledgements}
I would like to thank Professor Tsallis for making
valuable suggestions to improve this paper and A.K. Rajagopal for fruitful
conversations and perceptive comments.


\begin{thebibliography}{99}
\bibitem{Abe0} S.\ Abe, S.\ Martinez, F.\ Pennini and A.\ Plastino, Phys.
Lett. A (2001), in press [cond-mat/0011012], (2001).

\bibitem{Tsallis1} C. Tsallis, J. Stat. Phys. {\bf 52}, 479 (1988).

\bibitem{Tsallis4} C. Tsallis, Braz. J. Phys. {\bf 29}, 1 (1999). This is a
special issue on Tsallis thermostatistics and can be obtained from
http://www.sbf.if.usp.br/WWW\_pages/Journals /BJP/Vol129/Num1/index.htm.\\ 
See also http://tsallis.cat.cbpf.br/biblio.htm for a regulary updated
bibliography on the subject.

\bibitem{Zanette1} D.H.\ Zanette and P. A. Alemany, Phys.\ Rev.\ Lett. {\bf %
75}, 366 (1995).

\bibitem{Naudts1} J. Naudts, Dual description of nonextensive ensembles, to
appear in the Proc. of the ``{\em International Workshop on Classical and
Quantum Complexity and Nonextensive Thermodynamics}'' (Denton, Texas, 3-6
April 2000), eds. P. Grigolini, C. Tsallis and B.J. West,{\em \ Chaos,
Solitons and Fractals} (2001), in press.[cond-mat/9904070].

\bibitem{Abe4} S. Abe and A. K. Rajagopal, J.\ Phys.\ A {\bf 33}, 8733,
(2000), preprint [cond-mat/0002159].

\bibitem{Abe3} S.\ Abe, S.\ Martinez, F.\ Pennini, A.\ Plastino, Phys. Lett.
A (2001), in press [cond-mat/0006109], (2000).

\bibitem{Raja1} A. K. Rajagopal, Phys.\ Rev.\ Lett. {\bf 76}, 3469 (1996).

\bibitem{Raja2} A. K. Rajagopal, Braz. J.\ Phys {\bf 29}, 61 (1999).

\bibitem{Springer} S.\ Abe, Y.\ Okamoto (eds), {\em Nonextensive Statistical
Mechanics and its Applications}, Springer (2001).

\bibitem{Tsallis2} C.\ Tsallis, Phys.\ Rev.\ Lett. {\bf 75}, 3589 (1995).

\bibitem{Pennini} F. Pennini, A. R. Plastino and A. Plastino, Physica A {\bf %
258}, 446 (1998).

\bibitem{Gibbs1} W.\ Gibbs, {\em Elementary Principles in Statistical
Mechanics} (Yale University Press, New Haven, 1902).

\bibitem{Prato1} D.\ Prato and C.\ Tsallis, Phys.\ Rev.\ E {\bf 60}, 2398
(1999).

\bibitem{Plastino1} A. R. Plastino and A. Plastino, Phys. Lett. A {\bf 193},
140 (1994).

\bibitem{Abe1} S. Abe, Phys. Lett. A {\bf 263} (1999) 424; Erratum {\bf 267}
(2000) 456.

\bibitem{Abe2} S. Abe, Physica A {\bf 269} (1999) 403.

\bibitem{Rajagopal} A. K. Rajagopal and S. Abe, Phys. Rev. Lett. {\bf 83},
1711 (1999).

\bibitem{Ubriaco1} M.R.\ Ubriaco, Phys.\ Rev. E {\bf 57}, 179 (1998).

\bibitem{Ubriaco2} M.R.\ Ubriaco, Phys.\ Rev. E {\bf 60}, 165 (1999).

\bibitem{Raja3} E.K.\ Lenzi, R.S.\ Mendel and A. K. Rajagopal, , Physica A 
{\bf 286}, 503, [cond-mat/9904100], (2000).

\bibitem{Balian1} R.\ Balian, N.L.\ Balazs, Ann.\ Phys., (NY) {\bf 179}, 97
(1987).
\end{thebibliography}
\end{document}